\documentclass[fleqn,usenatbib]{mnras_mod}

\usepackage{txfonts}

\usepackage[T1]{fontenc}
\usepackage{graphicx}    % Including figure files
\usepackage{amssymb}    % Extra maths symbols
\usepackage{enumerate}
\usepackage{hyperref}

\newcommand{\aref}[1]{\hyperref[#1]{Appendix~\ref{#1}}}

\newcommand\blfootnote[1]{%
  \begingroup
  \renewcommand\thefootnote{}\footnote{#1}%
  \addtocounter{footnote}{-1}%
  \endgroup
}

\renewcommand{\footnoterule}{%
  \kern -3pt
  \hrule width 0.2\textwidth height 1pt
  \kern 2pt
}

\hypersetup{linkcolor=red}          % red equations
\title[Universal kinematic bimodality]{{\Large Article}\\ \vspace{-15pt} \line(1,0){500}\\Universal bimodality in kinematic morphology and the divergent pathways to galaxy quenching}

\author[B. Wang et al.]{
	Bitao Wang,$^{1,2}$
	Yingjie Peng,$^{1,2}$
	Michele Cappellari$^{3}$
	\\
}

\date{}

\pubyear{2024}

\begin{document}
\label{firstpage}
\pagerange{\pageref{firstpage}--\pageref{lastpage}}
\maketitle

%%%%%%%%%%%%%%%%% BODY OF PAPER %%%%%%%%%%%%%%%%%%

\textbf{The hierarchical structure formation of our Universe inherently involves violent and chaotic episodes of mass assembly such as galaxy mergers.
The level of bulk rotation of the collisionless stellar systems of galaxies reflects to what extent the galaxies, on the other hand, have assembled their stars during tranquil and ordered formation history, which fosters the growth of cohesively rotating structures.
Observationally, galaxy populations show a wide spectrum of morphology and shapes, with different levels of rotational support.
Despite the obvious variety and complexity, in this work we find that at a given stellar mass of galaxies, the distribution of the intrinsic spin parameter $\lambda_{R_{\rm e},\mathrm{intr}}$, i.e. the normalized specific angular momentum of stars, appears to be universally bimodal among galaxies in all star formation states and also in different environments.
This ubiquitous bimodality in kinematic morphology evolves systematically with star formation and is particularly apparent for transitional galaxies of intermediate star formation rates, indicating that star formation quenching is proceeding separately within two distinct kinematic populations dominated by cold discs and hot spheroids.
We show that the two populations also have contrasting recent star formation histories and metal enrichment histories, which reveal their divergent pathways to formation and quenching.
}

\blfootnote{$^{1}$Department of Astronomy, School of Physics, Peking University, Beijing 100871, China $^{2}$Kavli Institute for Astronomy and Astrophysics, Peking University, Beijing 100871, China $^{3}$Sub-department of Astrophysics, Department of Physics, University of Oxford, Denys Wilkinson Building, Keble Road, Oxford OX1 3RH, UK}

Physical processes involving angular momentum transfer, such as galaxy mergers and disc instability, can alter stellar orbits and eventually lead to star ensembles of different morphology \cite{1980MNRAS.193..189F, 1998MNRAS.295..319M, 2014MNRAS.438.1870D, 2015ARA&A..53...51S}.
Modern surveys of integral field spectroscopy \cite{2002MNRAS.329..513D, 2011MNRAS.413..813C, 2012A&A...538A...8S, 2015MNRAS.447.2857B, 2015ApJ...798....7B} provide spatially resolved kinematic maps for thousands of galaxies and have enabled statistical studies of angular momentum for galaxy populations across diverse environments.
Previous works \cite{2011MNRAS.414..888E, 2016ARA&A..54..597C, 2018MNRAS.477.4711G, 2020MNRAS.495.1958W, 2021MNRAS.505.3078V, 2022ApJ...937..117F} have discussed how the galaxy formation history might be related to the kinematic properties of these local populations.
However different conclusions have been drawn on whether the underlying distributions of various quantifications of galaxy rotational support can be considered bimodal, or whether there are more complex factors at play.

\begin{figure*}
	\includegraphics[width=0.98\textwidth]{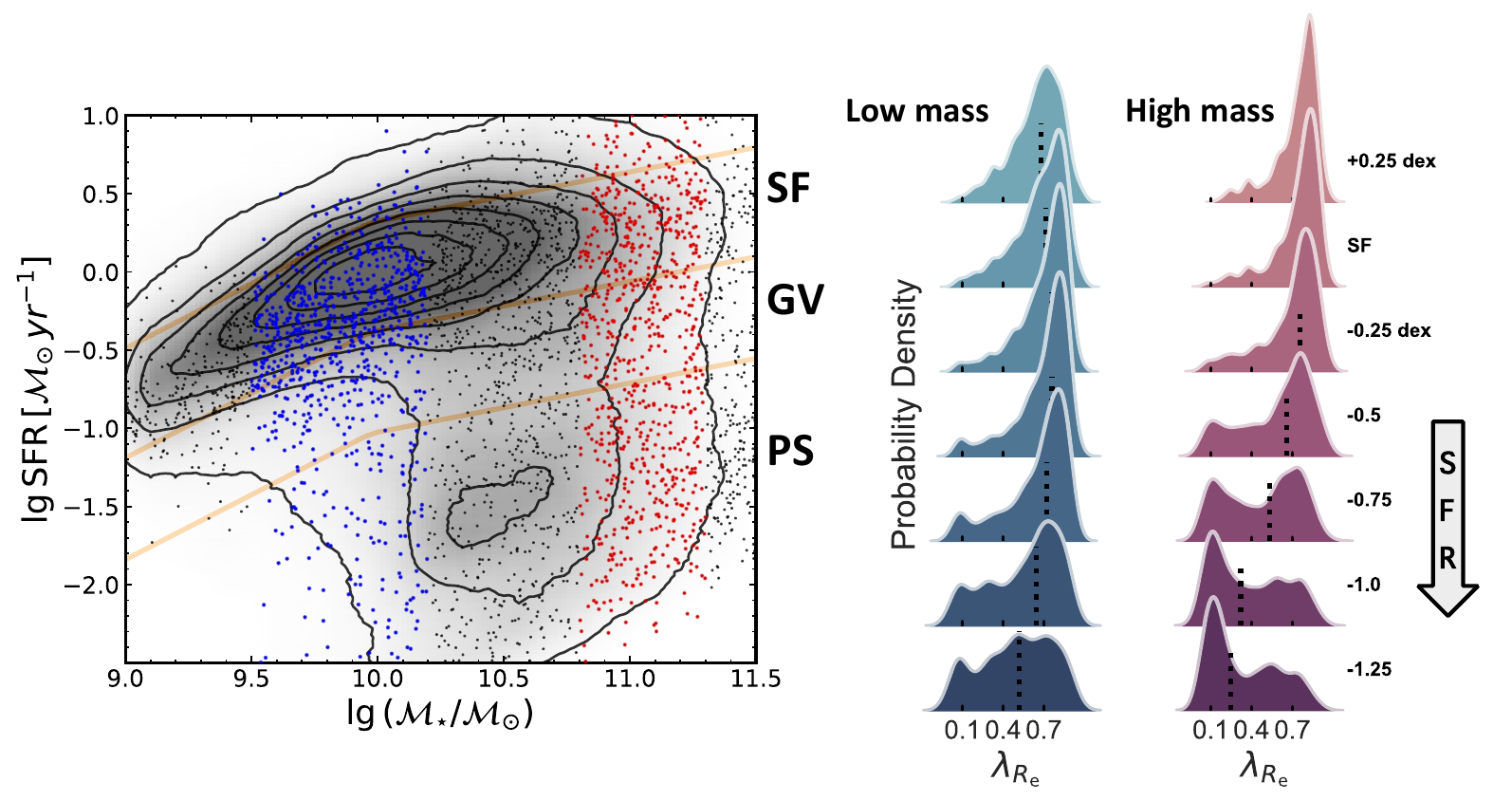}
	\caption{\textbf{A highlight of the bimodal $\lambda_{R_{\rm e}}$ variation with SFR among massive galaxies.}
	Left-hand panel: The SFR versus stellar mass diagram for MaNGA galaxies (dots) and a much larger sample of galaxies from the main SDSS spectroscopic survey (shaded contours).
	The three yellow lines mark the loci $+0.35$, $-0.35$, and $-1$ dex from the SFMS, and the SFMS is determined by the ridge of the upper contour, as a benchmark of normal star formation.
	Throughout this work, galaxies are divided into three star formation levels by the distance to the SFMS: SF, $-0.35<\Delta\,\mathrm{lg\, SFR}<0.35$; GV, $-1<\Delta\,\mathrm{lg\, SFR}<-0.35$; PS, $\Delta\,\mathrm{lg\, SFR}<-1$.
	Among MaNGA galaxies, blue and red dots show the positions of those in representative low-mass and high-mass bins respectively (blue: $10^{9.5}<\mathcal{M}_{\star}/\mathcal{M}_{\odot}<10^{10.2}$; red: $10^{10.8}<\mathcal{M}_{\star}/\mathcal{M}_{\odot}<10^{11.3}$).
	Right-hand panels: Two columns, coloured correspondingly, show the $\lambda_{R_{\rm e}}$ distribution as a function of SFR respectively for MaNGA galaxies in the two representative mass bins.
	Each panel shows the kernel estimated density distribution and the median value (black dotted line) of $\lambda_{R_{\rm e}}$ of galaxies in a certain $\Delta\,\mathrm{lg\, SFR}$ window of width 0.7 dex.
	From top to bottom the center of the window goes from high to low star formation in steps of 0.25 dex, with the window centers relative to the SFMS denoted on the rightmost.
	}
	\label{fig:bimo}
\end{figure*}

Taking a step further from our previous study \cite{2020MNRAS.495.1958W}, where we show that $\lambda_{R_{\rm e}}$ on average reduces with decreasing star formation rate (SFR), here we explore the detailed distributions of $\lambda_{R_{\rm e}}$ at given stellar mass and SFR.
The dots in the left panel of \autoref{fig:bimo} shows the galaxies studied in this work on the SFR versus stellar mass $\mathcal{M}_{\star}$ diagram.
These $\sim 3,000$ local galaxies, covering a wide range in morphology and environment, have available maps of stellar mean velocity and velocity dispersion from the Mapping Nearby Galaxies at Apache Point Observatory survey \cite{2015ApJ...798....7B} (MaNGA), as a part of the Data Release 15 of The Sloan Digital Sky Survey\cite{2019ApJS..240...23A} (SDSS). With quality control, the spin parameter $\lambda_{R_{\rm e}}$ is determined based on MaNGA kinematic maps \cite{2022MNRAS.511..139B}, and the stellar mass and SFR are derived via fitting the spectral energy distribution (SED) from ultraviolet (UV) to infrared (IR) wavelength \cite{2016ApJS..227....2S, 2018ApJ...859...11S}.

To quantify the star formation state in the broad context of local Universe, we
use $\sim 700,000$ galaxies in the same redshift range and with consistent mass and SFR measurements, which is a major part of the Main Galaxy Sample \cite{2002AJ....124.1810S} of SDSS.
The contours and gray shade illustrate the density distribution of these $\sim 700,000$ galaxies from which we define the ridge line (with volume correction) of the upper cloud of star-forming galaxies as the star formation main sequence (SFMS; see ref.\cite{2015ApJ...801L..29R}).
Galaxies around the SFMS are thought to be in dynamical equilibrium between gas inflow, star formation, and gas outflow \cite{2013ApJ...772..119L, 2014MNRAS.443.3643P}.
We define those with $-0.35<\Delta\,\mathrm{lg\, SFR}<0.35$ from the SFMS (between the upper two yellow lines) as normal star-forming galaxies (SF galaxies).
While galaxies below the lowest yellow line, which have SFR less than ten percent of star-forming galaxies on the SFMS ($\Delta\,\mathrm{lg\, SFR}<-1$), are quiescent and passive (PS galaxies).
Other galaxies with intermediate SFR ($-1<\Delta\,\mathrm{lg\, SFR}<-0.35$) are in the "valley" of the density distribution.
These galaxies generally also have intermediate colour index as compared to SF and PS galaxies, and are commonly defined as green valley (GV) galaxies.

In the right two columns of \autoref{fig:bimo}, we highlight how the normalized $\lambda_{R_{\rm e}}$ distribution varies with SFR among representative low- and high-mass  galaxies (blue: $10^{9.5}<\mathcal{M}_{\star}/\mathcal{M}_{\odot}<10^{10.2}$; red: $10^{10.8}<\mathcal{M}_{\star}/\mathcal{M}_{\odot}<10^{11.3}$).
From top to bottom, each shows the probability density distribution derived from Gaussian kernel estimation and the median (black dotted line) of $\lambda_{R_{\rm e}}$ for galaxies in sliding windows (0.7 dex wide and 0.25 dex step) of distance to the SFMS.
The positions of the windows with respect to the SFMS are marked on the rightmost.

For high-mass galaxies, from the SFMS to lower star formation levels, a twin peak at low $\lambda_{R_{\rm e}}$ gradually emerges while the high $\lambda_{R_{\rm e}}$ peak diminishes.
The change of the median value of $\lambda_{R_{\rm e}}$ is hence primarily due to the variation in the relative importance of the two peaks, rather than the move or broadening of individual peaks in the distribution.
Noteworthily, the median or average value of $\lambda_{R_{\rm e}}$ is far from being a good representative of the bimodal distribution among massive GV galaxies.
We have checked that this $\lambda_{R_{\rm e}}$ bimodality clearly exists even if we do not correct for atmospheric smearing and it is not due to any bimodality in the distribution of SFR or mass.
There is a hint for a similar separation among low-mass galaxies.
However, the growing tail at low $\lambda_{R_{\rm e}}$ toward low star formation levels only contains a small number of galaxies.
Previous evidence suggests that many of them host counter-rotating disks \cite{2022MNRAS.511..139B} and are not genuine triaxial slow rotators like the massive systems.
The nature of these seemingly slow-rotating low-mass systems also bears uncertainty resulted from lower signal-to-noise ratio (S/N) and spatial resolution compared to massive galaxies.

To remove the projection effect and determine the fractions of the underlying fast- and slow-rotating population, we build theoretical models and estimate the distributions of the intrinsic $\lambda_{R_{\rm e}}$ (i.e. $\lambda_{R_{\rm e}}$ observed edge-on).
The models we use are based on the tensor virial theorem that links together shape, ordered to random motion ratio, and velocity anisotropy for axisymmetric galaxies \cite{2005MNRAS.363..937B}.
The theoretical galaxy population constructed with the models are assumed to have an intrinsic ellipticity $\varepsilon_\mathrm{intr}$ distribution composed of two Gaussian components which stand for the fast- and slow-rotating population respectively.
The theoretical population are then projected onto the sky plane at random orientation, giving a projected $\lambda_{R}$ distribution to be compared with observations.
A simplification is implied in the above modelling.
The slow rotators of low $\lambda_{R,\mathrm{intr}}$ in real world have diverse shapes \cite{2007MNRAS.379..401E, 2011MNRAS.414..888E, 2018NatAs...2..483V}.
While some of them are mildly triaxial \cite{2014MNRAS.444.3340W} and some have not so small ellipticity due to the existence of counterrotating discs, here we treat them all as axisymmetric galaxies with single kinematic component, for them to be readily described by the tensor virial theorem.
This simplification does not affect the conclusion given our focus on the angular momentum of galaxies rather than their detailed shapes.
We get consistent results if we just model galaxies excluding the classic slow rotators (defined by equation 19 of ref. \cite{2016ARA&A..54..597C}) and integrate the observed $\lambda_{R_{\rm e}}$ distribution of classic slow rotators into the derived $\lambda_{R,\mathrm{intr}}$ distributions of other galaxies.

\begin{figure}
	\includegraphics[width=0.47\textwidth]{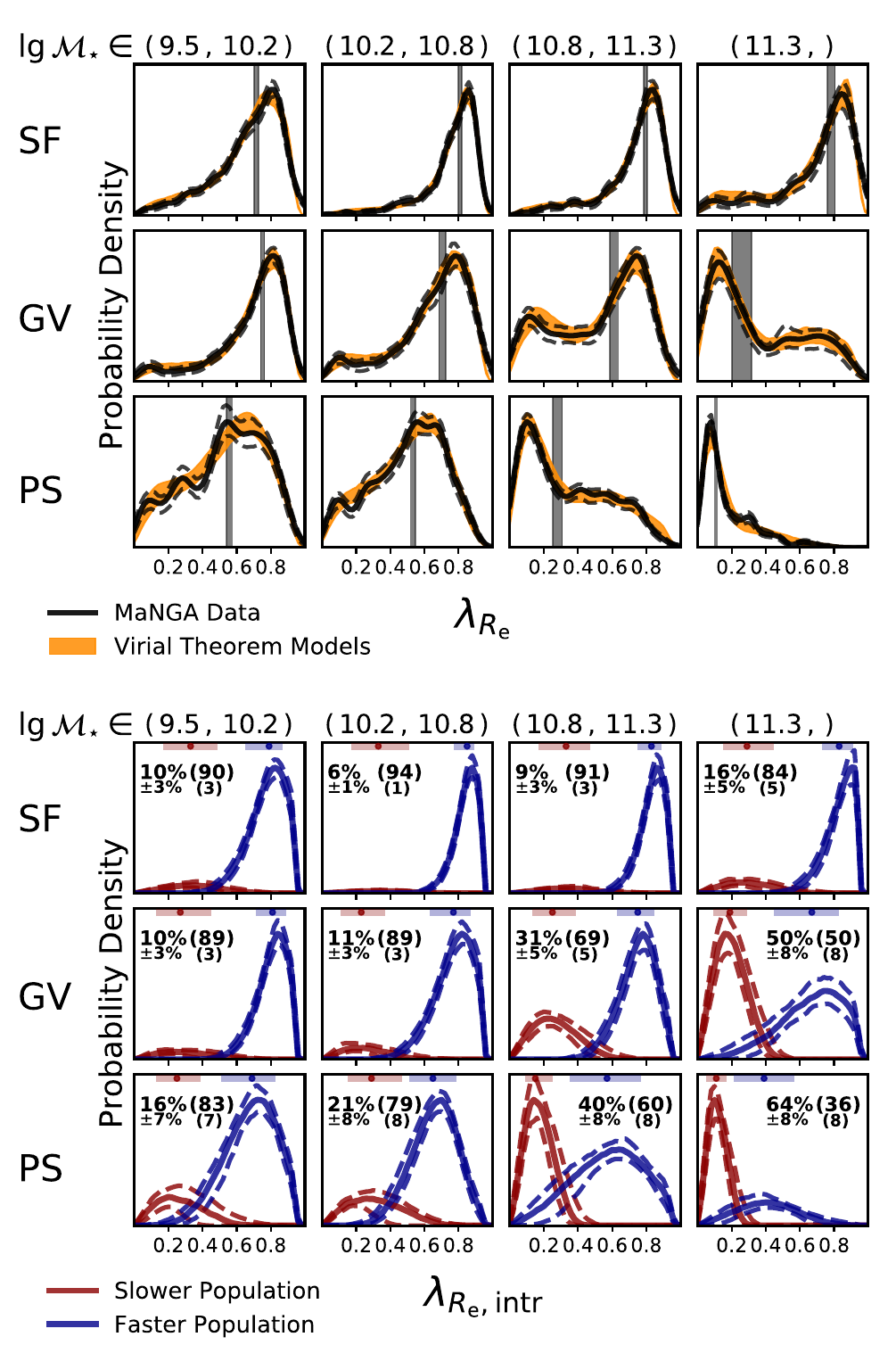}
	\caption{\textbf{The observed distributions of stellar spin $\lambda_{R_{\rm e}}$ and the corresponding modelled distributions of the inclination corrected intrinsic stellar spin $\lambda_{R_{\rm e},\mathrm{intr}}$.}
	Upper half: The black solid line in each panel is the observed $\lambda_{R_{\rm e}}$ density distribution of MaNGA galaxies in a certain bin of stellar mass and star formation level, with its uncertainty shown by the black dashed lines which illustrate the inner $68\%$ (in between $16\%$ and $84\%$ percentiles) of the density profiles of one hundred bootstrap samples.
	The range covered by the inner $68\%$ of all median $\lambda_{R_{\rm e}}$ of individual bootstrap samples is illustrated by a gray vertical band.
	While the orange band shows the inner $68\%$ of the density profiles of the best fitting models (models of two kinematic populations in virial equilibrium) for the MaNGA and bootstrap samples.
	As denoted, the stellar mass of each column increases rightward, and the star formation level of each row decreases downward from SF (star-forming) to GV (green valley), and to PS (passive).
	Lower half: The $\lambda_{R_{\rm e},\mathrm{intr}}$ distribution and its uncertainty of the best fitting models, for the two kinematic populations in the model with intrinsically faster (blue lines) and slower (red lines) rotation.
	In each panel, the numbers are fractions with error of the slow-rotating (fast-rotating) populations, and the dotted bars at the top edge show the median positions (the dot) and the inner $68\%$ of the two density distributions.
	}
	\label{fig:lam}
\end{figure}

The black solid lines in the upper half of panels of \autoref{fig:lam} show the observed $\lambda_{R_{\rm e}}$ probability density distribution in four mass bins (denoted at the top) for SF, GV, and PS galaxies respectively.
One hundred bootstrap samples are created and the inner $68\%$ (in between $16\%$ and $84\%$ percentiles) of their density distributions are shown by the black dashed lines to measure the uncertainties.
We then fit our two-component model to the observed and the bootstrap density distributions, and the inner $68\%$ of the best fitting models (determined via the least square method) are shown as the orange band.
The lower half of panels of \autoref{fig:lam} illustrate the corresponding $\lambda_{R_{\rm e},\mathrm{intr}}$ distributions of the best fitting two-component models, denoted with the relative percentage of the two distinct kinematic populations with intrinsically faster and slower rotation.

The estimated distributions of $\lambda_{R_{\rm e},\mathrm{intr}}$ in the lower panels of \autoref{fig:lam} do not differ significantly from the observed $\lambda_{R_{\rm e}}$ distributions above, indicating a weak projection effect overall.
By showing the bimodality variation as a function of star formation levels, the lower panels of \autoref{fig:lam} indicate that the two kinematic populations are most distinct with comparable importance when galaxies are massive and of intermediate star formation (GV and $\mathcal{M}_{\star}>10^{10.8}\mathcal{M}_{\odot}$).
The fraction of galaxy population with slower rotation generally increases with increasing stellar mass and decreasing SFR, ranging from $5\%$-$10\%$ among SF galaxies of relatively low mass to $65\%$ among most massive PS galaxies.
The slow-rotating population only make up a minority in a large mass range below $10^{11}\,\mathcal{M}_{\odot}$ (see also Fig.13 of ref. \cite{2018MNRAS.477.4711G} and Fig.3 of ref. \cite{2022ApJ...937..117F}).

The non-negligible fractions of the intrinsically slow-rotating population among SF galaxies indicate that the tails of the observed $\lambda_{R_{\rm e}}$ distributions of SF galaxies are not merely due to projection effect (i.e. not just face-on discs).
This can also be seen in Extended Data Fig.1 which shows the distributions on the $\lambda_{R_{\rm e}}$ versus $\varepsilon$ diagram for galaxy subsamples as in \autoref{fig:lam}.
Among SF galaxies in the upper row (represented by black dots), for those that have low values of apparent $\lambda_{R_{\rm e}}$ (e.g., $\lambda_{R_{\rm e}}<0.4$), some of them do have fairly round shapes with little $\varepsilon$ and are consistent with being face-on discs, i.e. those that are connected with relatively edge-on discs at $\lambda_{R_{\rm e}}\sim\varepsilon\sim0.8$ by the black dotted lines predicted by the tensor virial theorem.
But many more galaxies in the low-$\lambda_{R_{\rm e}}$ regime are too flattened to be face-on discs.
More importantly, if not decisively, in the following we will see that the slow-rotating population manifest drastically different stellar population properties from their fast-rotating counterparts.
This argues strongly against the possibility that the SF slow-rotating population are contaminated largely by face-on discs in which case there should be no significant difference in the stellar populations of SF slow- and fast-rotating galaxies.
Unlike the archetypical slow rotators among massive passive galaxies (predominantly occupying the lower left trapezoid area in Extended Data Fig.1), the SF slow-rotating population may generally possess residual disc structure (e.g., counterrotating discs seen from the kinematic maps), similar to the low-mass slow-rotating early-type galaxies studied in ref.\cite{2020A&A...635A.129K}.

Notably, with decreasing SFR, the lower panels of \autoref{fig:lam} show that the $\lambda_{R_{\rm e},\mathrm{intr}}$ distributions of the fast-rotating population broaden and move toward lower values, as shown by the dotted bars at the upper edge of panels which mark the position and width of the distributions.
This phenomenon is more obvious at higher stellar mass.
Massive slow-rotating populations further lose their residual angular momenta as they quench the star formation, and the oldest PS systems among them have spiked $\lambda_{R_{\rm e},\mathrm{intr}}$ distributions that are concentrated at values close to zero.
These trends of higher order terms can be also seen in the apparent $\lambda_{R_{\rm e}}$ distributions.

We find that the bimodality persists in diverse environment, quantified for example by the mass of dark matter halos from the catalogue\cite{2007ApJ...671..153Y} based on abundance matching, or by nearest neighbors defined number densities \cite{2019arXiv191005136G} that reflect more about the local environment.
With the dedicated group catalogue\cite{2019arXiv191005136G} for MaNGA galaxies, Extended Data Fig.2 shows the variation of bimodality between large (the left-hand side panels; the number of group members $\mathcal{N}\geq6$) and small galaxy groups (the right-hand side panels; $\mathcal{N}<6$), for galaxies in the same bins as in \autoref{fig:lam}.
$\mathcal{N}$ does not only indicate the galaxy richness in a group but is also a good indicator of the mass of dark matter halos with a Spearman rank correlation coefficient between the two $\sim0.9$.
The specific definition of large and small groups is made considering that the median galaxy groups in our sample host about six members.
Extended Data Fig.2 shows that the significant bimodal populations among massive GV galaxies appear to be even more distinct in larger galaxy groups.
For SF low-mass galaxies, intriguingly the fraction of the slow-rotating population in larger groups is nearly zero and is clearly lower than in smaller groups, which can also be seen from the observed $\lambda_{R_{\rm e}}$ distributions alone.

\begin{figure*}
	\includegraphics[width=0.98\textwidth]{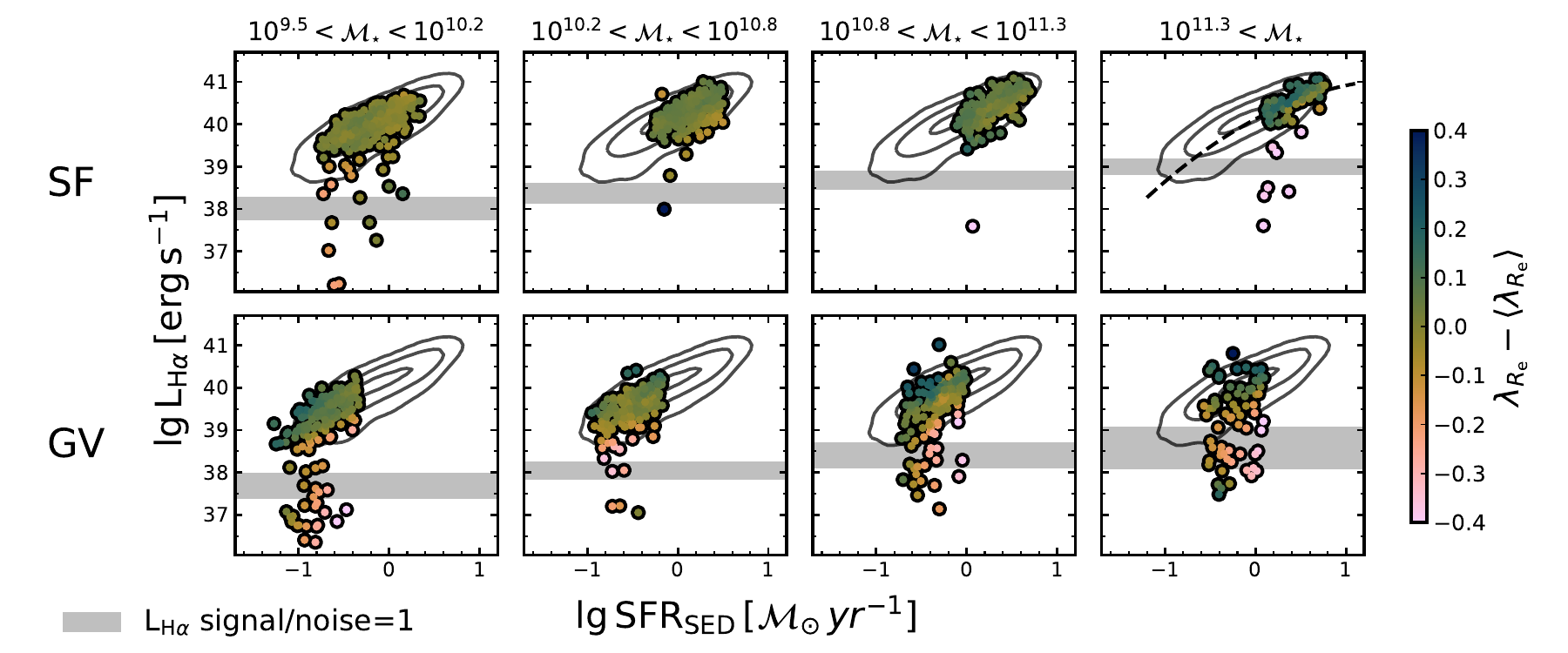}
	\caption{\textbf{The correlation between stellar angular momentum and recent star formation history.}
	Each panel shows, on logarithmic scale, the integrated luminosity of H$\alpha$ emission line (SFR over short timescale) versus SFR derived from UV, optical, and mid-IR SED (SFR over long timescale) for MaNGA galaxies in a certain bin of mass and star formation level (bins are the same as in \autoref{fig:lam} except that PS galaxies are not displayed due to their large errors on both axes).
	The gray horizontal band in each panel shows the noise level, represented by 1-$\sigma$ width of the distribution of integrated (over pixels within $1.5\,R_{\rm e}$) H$\alpha$ line noise.
	Data points are colour coded by the spin difference from the average at the mass and SFR of the galaxy, with LOESS smoothing applied to emphasize the trend that galaxies of lower spin tend to have lower H$\alpha$ luminosity.
	The scatter from this mean trend is presented in Extended Data Fig.3.
	The gray lines, same in every panel, are contours enclosing $90\%$, $68\%$, and $20\%$ of all galaxies in this figure, showing the general consistency between SFRs over short and long timescales.
	}
	\label{fig:halpha}
\end{figure*}

The drastically different kinematics of the two populations at the same mass and SFR implies at least two divergent evolution pathways.
We first probe their recent star formation histories (SFHs) using measurements that indicate SFRs respectively over short and long timescales.
The luminosity of H$\alpha$ emission line can trace the ionizing photons emitted by the most massive and short-lived stars, thus the SFR over recent $\sim 10\, \mathrm{Myr}$.
While the SFRs based on UV to mid-IR SED are averaged over $100\,\mathrm{Myr}$, a timescale close to the lifespan of young stars that emit mainly in the UV wavelength.
Galaxies with smooth SFHs such as gradual decline of SFR will show consistent values of H$\alpha$ luminosity and UV-based SFR, and those with fast-changing SFHs like burst or fast cessation of star formation can produce incompatible H$\alpha$ luminosity and UV-based SFR, which occurs when most massive stars in the young stellar populations have died.

The circles in \autoref{fig:halpha} show the MaNGA galaxies on H$\alpha$ luminosity versus SFR diagram, separately for SF (upper panels) and GV (lower panels) galaxies in four different mass bins.
Each galaxy is colour coded by $\lambda_{R_{\rm e}} - \langle\lambda_{R_{\rm e}}\rangle$, where $\langle\lambda_{R_{\rm e}}\rangle$ is the value averaged over galaxies of similar $\mathcal{M}_{\star}$ and SFR via the locally weighted regression method LOESS \cite{2013MNRAS.432.1862C}.
We have also implemented LOESS in each panel of \autoref{fig:halpha} to uncover the mean trend.
The H$\alpha$ is integrated within $1.5\,R_{\rm e}$ which is the observation coverage for the majority of MaNGA galaxies, regardless of the ionization type.
The results are consistent if we only include high S/N emissions with unambiguous star formation origin according to the BPT diagram \cite{1981PASP...93....5B, 2003MNRAS.346.1055K}, and take extinction into account.
The PS galaxies are not shown here because of their low H$\alpha$ luminosity and large uncertainties in SFRs.

The gray lines, marking the contours enclosing $90\%$, $68\%$, and $20\%$ of all galaxies studied in \autoref{fig:halpha}, show a clear correlation between H$\alpha$-based and UV-based SFRs.
This means sharp transition in the overall star formation of galaxies is not common at low redshift.
There are still several SF galaxies with low H$\alpha$ luminosity deviating the main trend, seemingly more at low and high mass end, and the fraction of deviating galaxies clearly increases in GV.
Noteworthily, galaxies on the main correlation denoted by the gray contours are mostly fast-rotating with positive $\lambda_{R_{\rm e}} - \langle\lambda_{R_{\rm e}}\rangle$, while those deviating from this correlation rotate more slowly with minus $\lambda_{R_{\rm e}} - \langle\lambda_{R_{\rm e}}\rangle$ values.
Extended Data Fig.3 confirms the result, under no LOESS smoothing, by directly showing $\lambda_{R_{\rm e}} - \langle\lambda_{R_{\rm e}}\rangle$ versus lg$\,\mathrm{L}_{\mathrm{H}\alpha}-\langle\mathrm{lg}\,\mathrm{L}_{\mathrm{H}\alpha}\rangle$ for each galaxy, where the latter quantifies the deviation of H$\alpha$ luminosity from the mean trend (the quadratic fit to all galaxies shown by the black dashed curve in the upper rightmost panel of \autoref{fig:halpha}).
The fraction of galaxies with smaller spin, or of the classic slow rotators (defined by equation 19 of ref.\cite{2016ARA&A..54..597C}) as marked by the red circles, generally increases when the H$\alpha$ emissions of galaxies are less luminous than the population average.
This trend indicates that fast cessation of star formation is more common in the population with slower rotation, consistent with previous works \cite{2014MNRAS.440..889S, 2018MNRAS.473.2679S}.

We further explore the SFHs of the fast- and slow-rotating population using stellar metallicity.
The stellar metallicity difference between star-forming and passive galaxies is an effective indicator to differentiate different quenching scenarios \cite{2015Natur.521..192P}.
Rapid removal of gas through strong outflows or stripping can produce a passive galaxy with similar stellar metallicity as its star-forming progenitor.
While the star formation of galaxies without gas supply, i.e. galaxies in the so-called starvation \cite{1980ApJ...237..692L}, reduces gradually and elevates the average stellar metallicity by the enriched interstellar medium.
This idea has been applied to large samples of galaxies in the local universe \cite{2020MNRAS.491.5406T}.

\begin{figure}
	\includegraphics[width=0.47\textwidth]{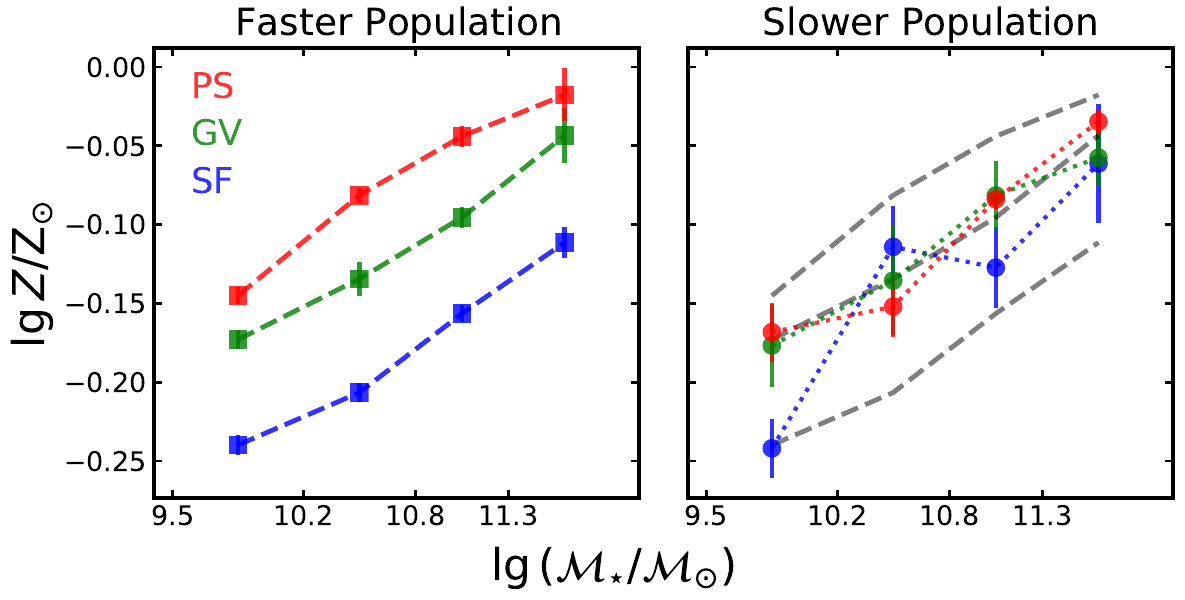}
	\caption{\textbf{The stellar metallicity versus mass relation for the two kinematic populations.}
	Left-hand panel: The relations for MaNGA galaxies with $\lambda_{R_{\rm e},\mathrm{intr}}>0.4$, divided into star formation levels of SF (star-forming), GV (green valley), and PS (passive) with different colours as labelled in the upper left corner.
	The mass bins are the same as in previous figures.
	Right-hand panel: The relations for those with $\lambda_{R_{\rm e},\mathrm{intr}}<0.4$, with the same star formation and mass binning as for on the left-hand panel.
	Additionally, the median relations in the left-hand panel, for the population that rotate faster, are also overlaid here for direct comparison.
	In both panels, each data point shows the median value with an error represented by the standard deviation of 1,000 bootstrap resamplings.
	Sizes of the samples used for statistics, from low to high mass shown in the following brackets: the faster population---SF (441/310/233/77), GV (183/160/127/35), PS (138/258/168/40); the slower population---SF (27/7/8/9), GV (18/21/38/33), PS (39/70/162/176).
	The results stay markedly consistent with the latest MaNGA data which more than double the size of each sample studied here.
	}
	\label{fig:metal}
\end{figure}

\autoref{fig:metal} shows the stellar metallicity versus stellar mass relation respectively for galaxies with faster (left panel) and slower (right panel) rotation, and separately for different star formation states as denoted in the upper left corner.
Each point with an error bar represents the median value and its uncertainty measured by the standard deviation of 1,000 bootstrap samples.
The stellar metallicity measurements normalized to the solar value $Z/\mathrm{Z}_{\odot}$ are taken from the P\textsc{ipe}3D value-added catalogue for MaNGA galaxies \cite{2016RMxAA..52..171S, 2018RMxAA..54..217S}.
These light-weighted measurements based on full spectrum fitting are all derived at one half-light radius $R_{\rm e}$ from line fitting to the metallicity radial profiles, so that they enable fair comparison between galaxies of different sizes while being representative of average metallicity over galactic scale \cite{2013A&A...554A..58S}.
Galaxies are divided into faster and slower population by their location relative to the $\lambda_{R_{\rm e}}$ versus $\varepsilon$ relation predicted by the tensor virial theorem for a galaxy with $\lambda_{R,\mathrm{intr}}=0.4$ (corresponding to $\varepsilon_\mathrm{intr}=0.525$ and $\delta=0.7\varepsilon_\mathrm{intr}=0.367$) projected at all inclinations.
The demarcation $\lambda_{R,\mathrm{intr}}=0.4$ is close to the distribution intersection in all but the lower right panels of the lower half of \autoref{fig:lam}.
We note that this demarcation is empirically based on the bimodal distributions and is notably higher than that for the classic slow rotators which are maximally spheroid-dominated systems.
Therefore, the slower population include both the classic slow rotators and spheroid-dominated galaxies yet with residual disc components.
The following result remains qualitatively the same if we study the radial profiles of metallicity, or use the latest \textsc{firefly} light-weighted metallicity measurements \cite{2017MNRAS.466.4731G, 2022MNRAS.513.5988N}, or adopt the classic fast and slow rotator classification.

Among the galaxy population with faster rotation, the stellar metallicity shows a significant increase with decreasing star formation at all masses.
By contrast, within errors the stellar metallicity of the galaxy population with slower rotation does not depend on star formation, except that the low-mass SF galaxies are relatively metal-poor at the given mass.
The PS slow-rotating populations are markedly less enriched than their fast-rotating counterparts.
Although minor mergers are able to develop metal-poor envelopes at the very outskirt of massive elliptical galaxies, they can hardly affect the main stellar body in the inner part \cite{2021ApJ...919..135D}, making the merger history an implausible reason for the observed metallicity difference.
Recent works \cite{2020ApJ...898...62B, 2022MNRAS.516.2971V} show that smaller galaxies with deeper gravitational potential wells may be enriched more efficiently.
We find that slow-rotating galaxies are generally more compact (smaller $R_{\rm e}$) than fast-rotating counterparts.
If potential does play a dominating role in the metal enrichment history (under debate; see ref. \cite{2023MNRAS.521.4173B} for example), then the deeper potential of the PS slow-rotating population yet with lower metallicity particularly elevates the importance of the quenching history in explaining their chemical property.
Together, these results fit well in the scenario of quenching histories \cite{2015Natur.521..192P} and suggest that the fast-rotating galaxies are primarily quenched through starvation, while the slow-rotating population through rapid gas removal.

The links between stellar angular momentum and SFHs might be natural consequences if discs grow quietly in quasi-equilibrium out of smooth and gradual gas inflows \cite{2020MNRAS.495L..42R, 2022MNRAS.515..213P} and spheroids build up in disruptive and chaotic processes, like significant mergers and violent disc instability \cite{2014MNRAS.444.3357N, 2015MNRAS.450.2327Z, 2017MNRAS.464.3850L, 2018MNRAS.480.4636S, 2022MNRAS.509.4372L}.
Star formation in fast-rotating galaxies can be quenched by preventive feedback \cite{2006MNRAS.365...11C, 2015ARA&A..53...51S} or quenched when gas is accreted with excessive angular momentum that hinders adequate radial inflows \cite{2020MNRAS.491L..51P}.
In both cases, the quenching proceeds slowly under starvation, with concurrent stellar metallicity enrichment.
While for the slow-rotating population, when massive dark matter halos are present \cite{2006MNRAS.368....2D, 2016MNRAS.457.2790T}, the violent processes they experience are able to halt the star formation in short time by gas depletion due to both star bursts and massive outflows \cite{2004ApJ...600..580G, 2023ApJ...950L..22W}, leaving stars less enriched in the end.

In this picture, it is expected that the fraction of slow-rotating massive galaxies increases sharply from SF to GV, as shown in \autoref{fig:lam}, given the presumably high quenching probability after violent events.
The highest peak of low $\lambda_{R_{\rm e},\mathrm{intr}}$ among most massive PS galaxies is observed and anticipated as these galaxies formed in the high peaks of the primordial density field with subsequent numerous mergers because of their gravitational dominance in the large scale structure \cite{2014ApJ...790L..24C}.
Over the relatively quiet formation history of disc populations, they could not always avoid disturbance and heating by external galaxy interactions and internal nonaxisymmetric structures, which explains the broadening and moving of the peak of high $\lambda_{R_{\rm e},\mathrm{intr}}$ with decreasing SFR.
However, compared to massive galaxies the weaker kinematic bimodality of low-mass galaxies implies more diverse formation histories \cite{2020A&A...635A.129K}, possibly due to their vulnerability to all kinds of effects.
For example, accretion of gas with misaligned angular momentum and its fueled star formation can slow down the average rotational velocity of stellar populations \cite{2022MNRAS.515..213P}.
Also, gravitational interactions with giant galaxies can heat the stars of low-mass satellites. These interactions, while not as devastating as mergers, may modestly reduce the rotational support of low-mass galaxies and make them kinematically intermediate systems.

To conclude, in this work we highlight the galaxy bimodality in kinematic morphology across different star formation levels in the local Universe, most significant for massive galaxies of intermediate star formation.
This ubiquitous kinematic bimodality, present also among galaxies of different masses and in diverse environments, pushes a step further from the acknowledged fast and slow rotator dichotomy \cite{2011MNRAS.414..888E, 2016ARA&A..54..597C, 2018MNRAS.477.4711G, 2020MNRAS.495.1958W}, and proves to be a general and fundamental property of galaxy populations.

Recent works \cite{2021MNRAS.505.3078V, 2022ApJ...937..117F} suggest that a simple bimodal description of galaxy stellar kinematics using two Gaussian components is not applicable for all galaxies and galaxy populations at certain stellar masses.
Our results suggest that this can be understood given that the fast- and slow-rotating modes of the $\lambda_{R_{\rm e},\mathrm{intr}}$ Gaussian components systematically vary with SFR.
Most notably the old and passive fast-rotating population are dynamically hotter and have less angular momentum than their young and star-forming counterparts, blurring the bimodality when galaxies of comparable mass but different SFR are viewed together.

It has long been suggested that the transitional galaxies in the green valley are typically disc plus spheroid composites with intermediate spheroid/total mass ratio, like the Sombrero Galaxy.
The suggestion comes from the well established empirical correlation between morphology and star formation level of galaxies \cite{2003MNRAS.341...54K, 2004ApJ...600..681B, 2009ApJ...699..105C, 2019MNRAS.485..666B}, where transitional galaxies have intermediate morphology on average.
Nevertheless, our results show that the average morphology, in particular for the massive transitional population, is not a valid representation of the underlying galaxy populations.
The kinematic bimodality indicates that the galaxies are closer to being either discs or spheroids than being composite systems in between.
The associated distinctions in metal enrichment and recent SFHs provide further evidence for the existence of two kinematic populations at certain mass and SFR, respectively with quiet and violent formation and quenching histories.

Quenching in the local Universe is thus rarely happening in galaxies comprising a slow-rotating bulge and a fast-rotating disc of comparable mass.
As opposed to previous belief, apart from the extremely massive galaxies the quenching of star formation at low redshifts are proceeding mostly in disc galaxies that are kinematically akin to common blue galaxies of vigorous star formation.
Toward the high mass end, the sharply growing prevalence of quenching via the slow-rotating channel is reminiscent of the significant increase of {\it ex situ} mass (i.e. from mergers) fraction with galaxy mass \cite{2019MNRAS.487.5416T}.
After a galaxy is quenched, it loses the ability to maintain or rebuild a disc via star formation, and hence steps upon a one-way trip to lower rotation unless it absorbs a large amount of orbital angular momentum from mergers of specific configurations.

Massive galaxies are luminous representatives of the overall cosmic structure.
Their unambiguous bimodal kinematic morphology, and the contrasting quenching histories associated with the two kinematic modes, imply the existence of divergent evolution pathways in the formation of the cosmos.
State-of-the-art simulations suggest a wide range of factors that together decide the stellar kinematics of galaxies, including halo spin, properties of mergers (e.g., orbital configurations, wetness, and mass ratios), and tidal interactions with environment \cite{2018MNRAS.476.4327L, 2020MNRAS.496.2673J, 2021MNRAS.507.3301Z}.
But none of these properties appears to be bimodal, nor have simulations directly shown such bimodal kinematics among massive galaxies \cite{2020MNRAS.498.4386P, 2021MNRAS.505.3078V}.
Future large surveys of integral field spectroscopy at higher redshifts may provide key clues by showing the temporal evolution of the kinematic bimodality, which carries indispensable information about how galaxies were formed.

\section*{Methods}\label{sec:method}

\begingroup
    \fontsize{8.pt}{9pt}\selectfont

		\textbf{The galaxy sample.} We take integral field spectroscopy data from the MaNGA survey, the part in the publicly available SDSS Data Release 15 \cite{2019ApJS..240...23A}.
		This data release includes 4,597 unique galaxies in the redshift range $0.01<z<0.15$ observed via hexagonal integral field units with angular resolution FWHM of 2.5 arcsec and effective diameters ranging from 12 to 32 arcsec \cite{2015AJ....149...77D}.
		The spectra cover from 360 to 1030 nm with median instrument broadening $\sigma_{\mathrm{inst}}\,\sim\,72\,\mathrm{km}\,s^{-1}$ (ref. \cite{2016AJ....152...83L}) and typical spectral resolution $R \sim 2000$.

		Our data are included in the three major subsamples \cite{2017AJ....154...86W} of MaNGA survey.
		The Primary and Secondary subsample are selected to ensure spatial coverage respectively out to 1.5 and 2.5 half-light radii $R_{\rm e}$, with a flat mass distribution.
		The Colour-Enhanced subsample is designed to increase the number of galaxies in the low-density regions of colour–magnitude diagram by extending the redshift limits of the Primary subsample in appropriate colour bins, so that the Primary plus the Colour-Enhanced subsamples smoothly cover the colour-magnitude diagram.
		No cuts are applied to galaxy colour, morphology, or environment so that MaNGA galaxies are fully representative of galaxy population in the local universe.

		We apply the same data quality control as in ref. \cite{2018MNRAS.477.4711G}, to remove data that are flagged bad or show signs of being problematic, and to exclude galaxies too small as compared with the MaNGA synthesized beam.
		Visually checking the SDSS optical images, galaxies in close pairs are excluded when their stellar light is highly blended with the half-light ellipses hard to define.
		Overlapping merging galaxies in chaotic morphology, i.e. significant onging mergers with low mass difference, are not included as they are far from being dynamically relaxed and the stellar kinematics is overwhelmed by the orbital motion between the galaxies rather than the internal stellar motion.
		Our sample captures the post-merger phases when galaxies have roughly resumed dynamical equilibrium.
		The above control discards 12\% of all galaxies and is meant for robust quantifications of internal stellar kinematics.
		After cross matching with the GALEX-SDSS-WISE Legacy Catalogue \cite{2016ApJS..227....2S,2018ApJ...859...11S} of total stellar mass and SFR for SDSS galaxies, finally we reach a sample of 3,279 galaxies among which 2,981 galaxies have stellar mass above $10^{9.5}\,\mathcal{M}_{\odot}$.

		In the consensus cosmology, mergers play an important role in galaxy evolution \cite{1978MNRAS.183..341W}.
		Removing the currently merging galaxies, which only make up a small fraction 3\%, certainly does not mean that the population studied have not been affected by mergers in the past.
		Supplementary Fig.1 shows the asymmetry of galaxy shapes under 180-degree rotation \cite{2016MNRAS.456.3032P} for fast- and slow-rotating galaxies in different star formation states.
		Across a large range of stellar masses, the shape asymmetry remains consistently at low levels (comparable to the median values of galaxies with regular morphologies presented by the panels (b) of Figure 11 in ref. \cite{2016MNRAS.456.3032P}), suggesting that galaxies in our sample that are severely out of dynamical equilibrium are rare, a consequence expected for our quality control.
		But inspection of optical images \cite{2022MNRAS.512.2222V} reveals the presence of tidal debris (e.g., shells, streams, tails) among a significant fraction of these galaxies, and indicates the occurrence of merger events in the near past \cite{2008ApJ...689..936J}.
		12\% of the star-forming fast-rotating galaxies have visually identified tidal features while this number almost doubles (21\%) for star-forming slow-rotating galaxies, for which we indeed expect more mergers.
		We note that the above fractions are only lower limits given the finite depth of the imaging data.\\[4pt]

		\noindent\textbf{The spin parameter $\lambda_{R_{\rm e}}$.} We use $\lambda_{R_{\rm e}}$ to quantify the level of rotational support of galaxies.
		The stellar kinematics can better reflect the intrinsic morphology of galaxies than projected light distributions \cite{2016ARA&A..54..597C}, so that the quantification $\lambda_{R_{\rm e}}$ is also referred to as a parameter of kinematic morphology.

		The measurements of $\lambda_{R_{\rm e}}$ for MaNGA galaxies are taken from ref. \cite{2022MNRAS.511..139B} with the methods detailed in ref.\cite{2018MNRAS.477.4711G}.
		The maps of stellar kinematics in the MAPS files are produced by the Data Analysis Pipeline of MaNGA \cite{2019AJ....158..231W}.
		The Data Analysis Pipeline first Voronoi bins \cite{2003MNRAS.342..345C} the data cubes to achieve a minimum signal-to-noise ratio of $\sim10$ per spectral bin of width $70\,\mathrm{km}\,s^{-1}$.
		Then the Penalised Pixel-Fitting method \cite{2017MNRAS.466..798C} is applied to extract the line-of-sight velocity distribution by fitting a set of 49 families of stellar spectra from the MILES stellar library \cite{2006MNRAS.371..703S, 2011A&A...532A..95F} to the observed absorption-line spectra, from where one can characterize the stellar kinematics by the mean stellar velocity V and velocity dispersion $\sigma$.
		$\lambda_{R_{\rm e}}$ is calculated using equation 5 and 6 of ref.\cite{2007MNRAS.379..401E}:
		\begin{equation}
		\lambda_{R}\,\equiv \, \frac{\left< R\,|V| \right>}{\left< R\, \sqrt{V^2+\sigma ^2} \right>}\,=\,\frac{\sum_{n=1}^{N} F_n R_n |V_n|}{\sum_{n=1}^{N} F_n R_n \sqrt{V_n^2+\sigma _n ^2}}
		\end{equation}
		where $F_n$, $V_n$, and $\sigma _n$ are the flux, mean velocity and velocity dispersion of the nth pixel.
		The summation is performed over the N pixels within radius R, and specifically for this work within the elliptical half-light radius $R_{\rm e}$ for $\lambda_{R_{\rm e}}$.

		Seeing corrections for the observed $\lambda_{R_{\rm e}}$ have been applied using analytic functions \cite{2018MNRAS.477.4711G}.
		Briefly, the effect of atmospheric smearing is quantified by measuring $\lambda_{R_{\rm e}}$ of realistic models of galaxy kinematics \cite{2008MNRAS.390...71C} convolved with a Gaussian point spread function of varying widths.
		An independent test \cite{2019MNRAS.483..249H} shows that the correction is effective and leaves little systematics.
		About 3\% of our galaxies have $\sigma_\mathrm{PSF}/R_{\rm e}^\mathrm{maj}>0.5$ which get large correction for smearing.
		The correction is only effective for and applied to galaxies with regular, hourglass-like rotation in their velocity field.
		Nonregular rotator galaxies are mostly slow-rotating with median $\lambda_{R_{\rm e}}=0.07$ and their underestimation of $\lambda_{R_{\rm e}}$ due to smearing is generally at the level of 0.1 \cite{2020MNRAS.497.2018H}. \\[4pt]

		\noindent\textbf{Ellipticity $\varepsilon$.}
		With the predictions from the tensor virial theorem, we use the observed ellipticity $\varepsilon$ of galaxy shape along with the measured $\lambda_{R_{\rm e}}$ to broadly classify galaxies into intrinsically flat and round objects (in \autoref{fig:metal}).
		The ellipticity, catalogued in ref. \cite{2022MNRAS.511..139B}, is measured by modelling the SDSS $r$-band photometry via the Multi-Gaussian Expansion method \cite{1994A&A...285..723E, 2002MNRAS.333..400C}.

		The ellipticity is calculated inside the model half-light isophote from the second moments of light distribution:
		\begin{equation}
		(1-\varepsilon)^2\,=\,q^{\prime 2}\,=\,\frac{\left< y^2 \right>}{\left< x^2 \right>}\,=\,\frac{\sum_{k=1}^{P} F_k y_k^2}{\sum_{k=1}^{P} F_k x_k^2}
		\end{equation}
		where $F_k$ and $(x_k,y_k)$ are the flux and coordinate of the $k$th pixel.\\[4pt]

		\noindent\textbf{Kinematic models.}
		We build kinematic models of axisymmetric galaxies via the tensor virial theorem \cite{2005MNRAS.363..937B}, which predicts the ordered to random motion ratio $(V/\sigma)_\mathrm{intr}$ with given intrinsic shape $\varepsilon_\mathrm{intr}$ and velocity anisotropy $\delta$.
		The formalism is summarized by eqs. 14--15 of ref. \cite{2016ARA&A..54..597C}.
		The predicted $(V/\sigma)_\mathrm{intr}$ is then translated into $\lambda_{R,\mathrm{intr}}$ using the tight relation derived from two-integral Jeans models \cite{2011MNRAS.414..888E}.

		We assume for our theoretical galaxy population that their $\varepsilon_\mathrm{intr}$ distribution, in the range (0,1), is composed of two gaussians truncated at zero and one.
		With the sampled $\varepsilon_\mathrm{intr}$ from the two-Gaussian distribution, we assign each model galaxy the velocity anisotropy $\delta$ randomly sampled from a uniform distribution in the empirically-motivated range $[0, 0.7\varepsilon_\mathrm{intr}]$ \cite{2007MNRAS.379..418C, 2021MNRAS.500L..27W}.
		The $\lambda_{R,\mathrm{intr}}$ can then be calculated for each galaxy in the model galaxy population, which are later projected onto the sky plane at random orientation using eqs. 16 of ref. \cite{2016ARA&A..54..597C}.
		We finally arrive at the projected $\lambda_{R}$ distribution of the model galaxy population to be compared with observations.

		When fitted to the data, the five free parameters of our theoretical galaxy population are the positions (2 parameters), standard deviations (2 parameters), and fractions (1 parameter) of the two Gaussian components in the assumed $\varepsilon_\mathrm{intr}$ probability density distribution.\\[4pt]

		\noindent\textbf{Star formation rate and stellar mass.} These two measurements are provided by the version X2 of the GALEX-SDSS-WISE Legacy Catalogue \cite{2016ApJS..227....2S,2018ApJ...859...11S} (GSWLC-X2) for SDSS galaxies in redshift range $0.01<z<0.3$ within the footprint of GALEX All-sky Imaging survey  \cite{2005ApJ...619L...1M}.
		Using the state-of-the-art modelling technique \cite{2009A&A...507.1793N, 2019A&A...622A.103B} (CIGALE), stellar mass and SFR are derived by fitting the SED of two GALEX UV flux, five SDSS optical plus near-IR flux, and one mid-IR flux (22 microns or 12 microns when the former is not available) from WISE \cite{2010AJ....140.1868W}.
		A Chabrier\cite{2003PASP..115..763C} initial mass function and a flat WMAP7 cosmology ($H_0$ = $70\ \mathrm{km}\,\mathrm{s}^{-1}\,\mathrm{Mpc}^{-1}$, $\Omega_\mathrm{m}$ = 0.27) are assumed.
		There are deeper GALEX UV observations nested in the All-sky Imaging footprint, and GSWLC-X2 makes use of the deepest available UV image for each galaxy.
		Note that GSWLC masses are systematically smaller than the dynamically-calibrated masses \cite{2013MNRAS.432.1709C} by $\sim0.4$ dex.
		This means that the characteristic mass $\mathcal{M}_\mathrm{crit}\approx10^{11.3}$ (ref. \cite{2016ARA&A..54..597C}) where slow rotators start dominating, appears at $\mathcal{M}_\mathrm{crit}\approx10^{10.9}$ in our sample.\\[4pt]

		\noindent\textbf{Stellar metallicity.} From the MaNGA-P\textsc{ipe}3D value-added catalogue we retrieve the light-weighted stellar metallicity measurements at one $R_{\rm e}$.
		The IA-UNAM MaNGA team created the catalogue by analyzing the MaNGA data cubes through the P\textsc{ipe}3D pipeline \cite{2016RMxAA..52..171S, 2016RMxAA..52...21S}, which fits the stellar continuum with stellar population models and measures the nebular emission lines of galaxies.
		A Salpeter \cite{1955ApJ...121..161S} initial mass function and a standard $\Lambda$CDM cosmology ($H_0$ = $73\ \mathrm{km}\,\mathrm{s}^{-1}\,\mathrm{Mpc}^{-1}$, $\Omega_\mathrm{m}$ = 0.3) are assumed.
		P\textsc{ipe}3D has been extensively tested against other spectrum fitting tools, and widely used in many other integral field spectroscopy surveys.

		The stellar metallicity values at one $R_{\rm e}$ are derived according to the linear regression of metallicity radial profiles and have been shown to match well the average metallicity across the optical extension of galaxies while with less error \cite{2013A&A...554A..58S}.
		We find consistent results by using mass-weighted metallicity or analyzing metallicity profiles.\\[4pt]

		\noindent\textbf{Data availability.} The MaNGA DR15 MAPS files of spatially resolved emission line properties and stellar kinematics are archived at https://dr15.sdss.org/sas/dr15/manga/spectro/analysis/v2\_4\_3/2.2.1/VOR10-GAU-MILESHC/.
		The measured $\lambda_{R_{\rm e}}$ and $\varepsilon$ for MaNGA DR15 galaxies are catalogued in ref.\cite{2022MNRAS.511..139B}, with the update to cover full MaNGA (DR17) recently published \cite{2023ApJ...950L..22W}.
		Stellar mass and SFR of the MaNGA galaxies are taken from the GSWLC-X2 catalogue \cite{2016ApJS..227....2S,2018ApJ...859...11S} available on https://salims.pages.iu.edu/gswlc/.
		The stellar metallicity measurements \cite{2016RMxAA..52..171S, 2018RMxAA..54..217S} that we use are publicly available on https://www.sdss.org/dr15/manga/manga-data/manga-pipe3d-value-added-catalog/.
		The shape asymmetry quantifications \cite{2016MNRAS.456.3032P} are taken from https://www.sdss4.org/dr17/data\_access/value-added-catalogs/?vac\_id=pawlikmorph-catalog-of-galaxy-morphologies.
		The visual identifications of tidal debris \cite{2022MNRAS.512.2222V} are retrieved from the catalogue at https://www.sdss4.org/dr17/data\_access/value-added-catalogs/?vac\_id=manga-visual-morphologies-from-sdss-and-desi-images.\\[4pt]

		\noindent\textbf{Code availability.} We use the Python package \textsc{loess} v2.0.11 available from https://pypi.org/project/loess/ to apply the locally weighted regression method LOESS \cite{2013MNRAS.432.1862C}.
		The \textsc{MgeFit} Python software package used for multi-Gaussian expansion \cite{1994A&A...285..723E, 2002MNRAS.333..400C} of galaxy surface brightness is available at https://pypi.org/project/mgefit/.
		The isophotal contour containing half the total luminosity of the multi-Gaussian expansion model is determined via the routine \textsc{mge\_half\_light\_isophote}, which is included in the \textsc{JamPy} Python software package \cite{2008MNRAS.390...71C} available at https://pypi.org/project/jampy/.

\endgroup

\section*{Acknowledgements}

We are grateful for the constructive comments by the anonymous referees and the extensive and encouraging discussions with them.
We thank very much our Editor Morgan Hollis for his comments and help, which lent us tremendous support over the reviewing process.
Y.P. and B.W. acknowledge National Science Foundation of China (NSFC) Grant No. 12125301, 12192220, 12192222, and the science research grants from the China Manned Space Project with No. CMS-CSST-2021-A07.

Funding for the Sloan Digital Sky Survey IV has been provided by the Alfred P. Sloan Foundation, the U.S. Department of Energy Office of Science, and the Participating Institutions. SDSS acknowledges support and resources from the Center for High- Performance Computing at the University of Utah. The SDSS website is \url{www.sdss.org}.

SDSS-IV is managed by the Astrophysical Research Consortium for the
Participating Institutions of the SDSS Collaboration including the
Brazilian Participation Group, the Carnegie Institution for Science,
Carnegie Mellon University, the Chilean Participation Group, the French Participation Group, Harvard-Smithsonian Center for Astrophysics,
Instituto de Astrof\'isica de Canarias, The Johns Hopkins University, Kavli Institute for the Physics and Mathematics of the Universe (IPMU) /
University of Tokyo, the Korean Participation Group, Lawrence Berkeley National Laboratory,
Leibniz Institut f\"ur Astrophysik Potsdam (AIP),
Max-Planck-Institut f\"ur Astronomie (MPIA Heidelberg),
Max-Planck-Institut f\"ur Astrophysik (MPA Garching),
Max-Planck-Institut f\"ur Extraterrestrische Physik (MPE),
National Astronomical Observatories of China, New Mexico State University,
New York University, University of Notre Dame,
Observat\'ario Nacional / MCTI, The Ohio State University,
Pennsylvania State University, Shanghai Astronomical Observatory,
United Kingdom Participation Group,
Universidad Nacional Aut\'onoma de M\'exico, University of Arizona,
University of Colorado Boulder, University of Oxford, University of Portsmouth,
University of Utah, University of Virginia, University of Washington, University of Wisconsin,
Vanderbilt University, and Yale University.

\section*{Author contributions}
B.W., Y.P. and M.C. all contributed extensively and significantly to the work presented in this paper, including the analyses, presentation, and interpretation of the data.
B.W. finished writing the first version of manuscript, after which all authors worked closely on revisions.

\section*{Competing interests}
The authors declare no competing interests.



%%%%%%%%%%%%%%%%%%%%%%%%%%%%%%%%%%%%%%%%%%%%%%%%%%

\vspace{+0.5cm}

{\bf Correspondence and requests for materials} should be addressed to B.W. (bt-wang@pku.edu.cn) and Y.P. (yjpeng@pku.edu.cn).

% \section*{Additional information}

%%%%%%%%%%%%%%%% APPENDICES %%%%%%%%%%%%%%%%%%%%%

\appendix

\begin{figure*}
	\includegraphics[width=0.98\textwidth]{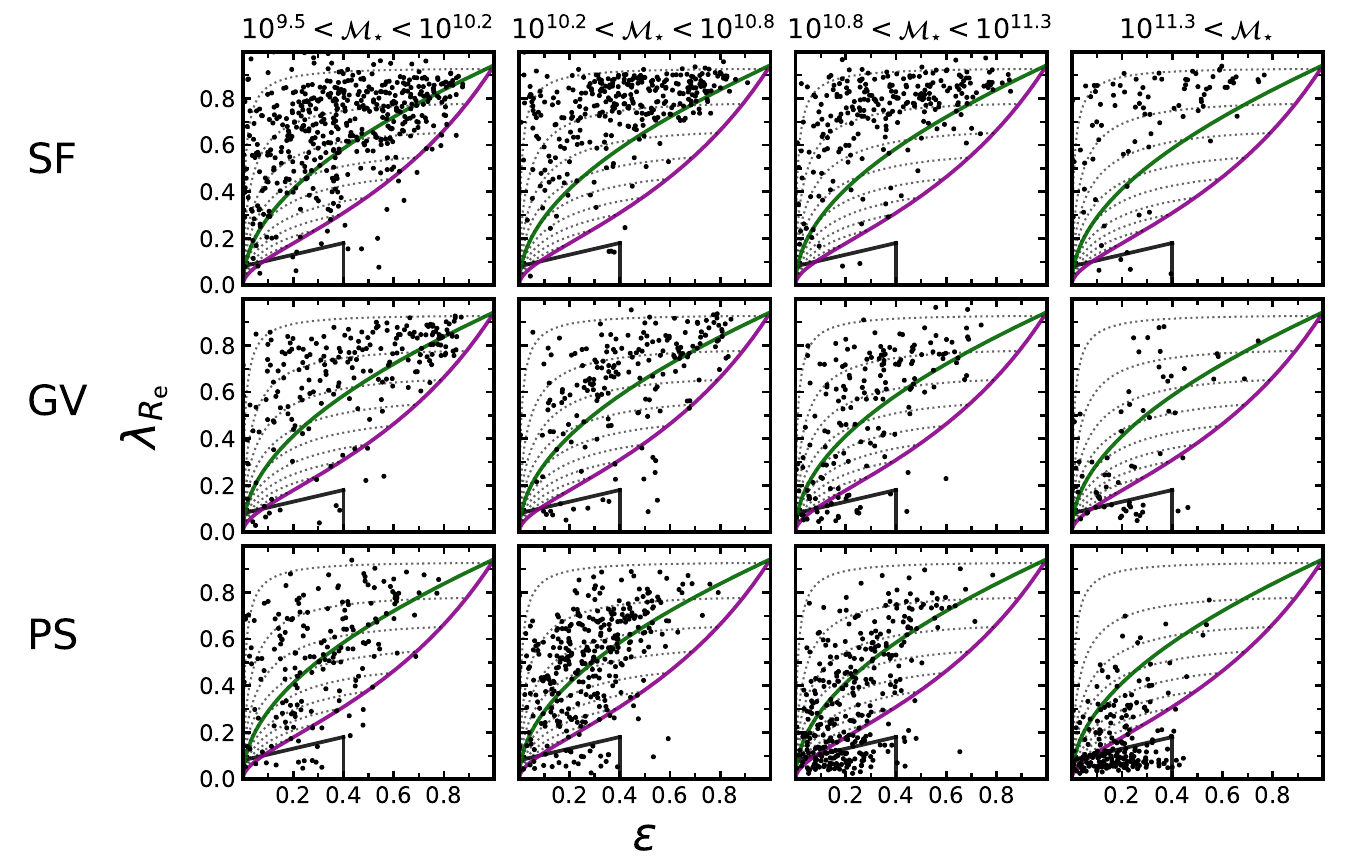}
	\caption{\textbf{Galaxy distributions on the $\lambda_{R_{\rm e}}$ versus $\varepsilon$ diagram.}
	The stellar mass and star formation bins are the same as in \autoref{fig:lam}.
	Black dots represent the observed MaNGA galaxies.
	The solid magenta line shows the locus of edge-on oblate rotators of varying intrinsic ellipticity $\varepsilon_\mathrm{intr}$, at the velocity anisotropy $\delta=0.7\varepsilon_\mathrm{intr}$, predicted by the tensor virial theorem.
	Starting from the magenta line, the black dotted lines show how the observed $\lambda_{R_{\rm e}}$ and $\varepsilon$ drop with decreasing inclination.
	While the magenta line assumes maximally anisotropic velocity dispersion based on empirical and theoretical evidence \citep{2007MNRAS.379..418C, 2021MNRAS.500L..27W}, the green line shows the locus of isotropic velocity dispersion.
	The classic slow rotators are defined by the trapezoidal area at the lower left.
	}
	\label{fig:le}
\end{figure*}

\begin{figure*}
	\includegraphics[width=0.47\textwidth]{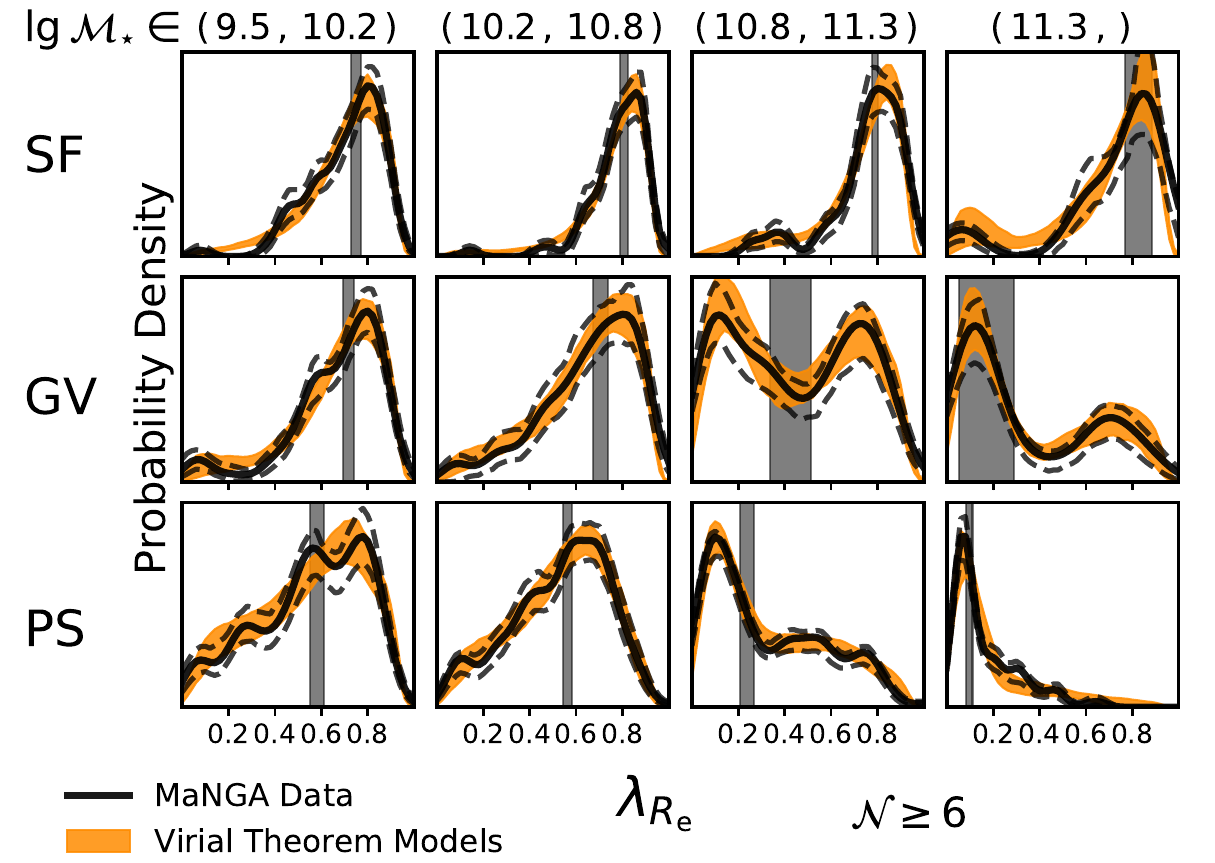}
	\includegraphics[width=0.47\textwidth]{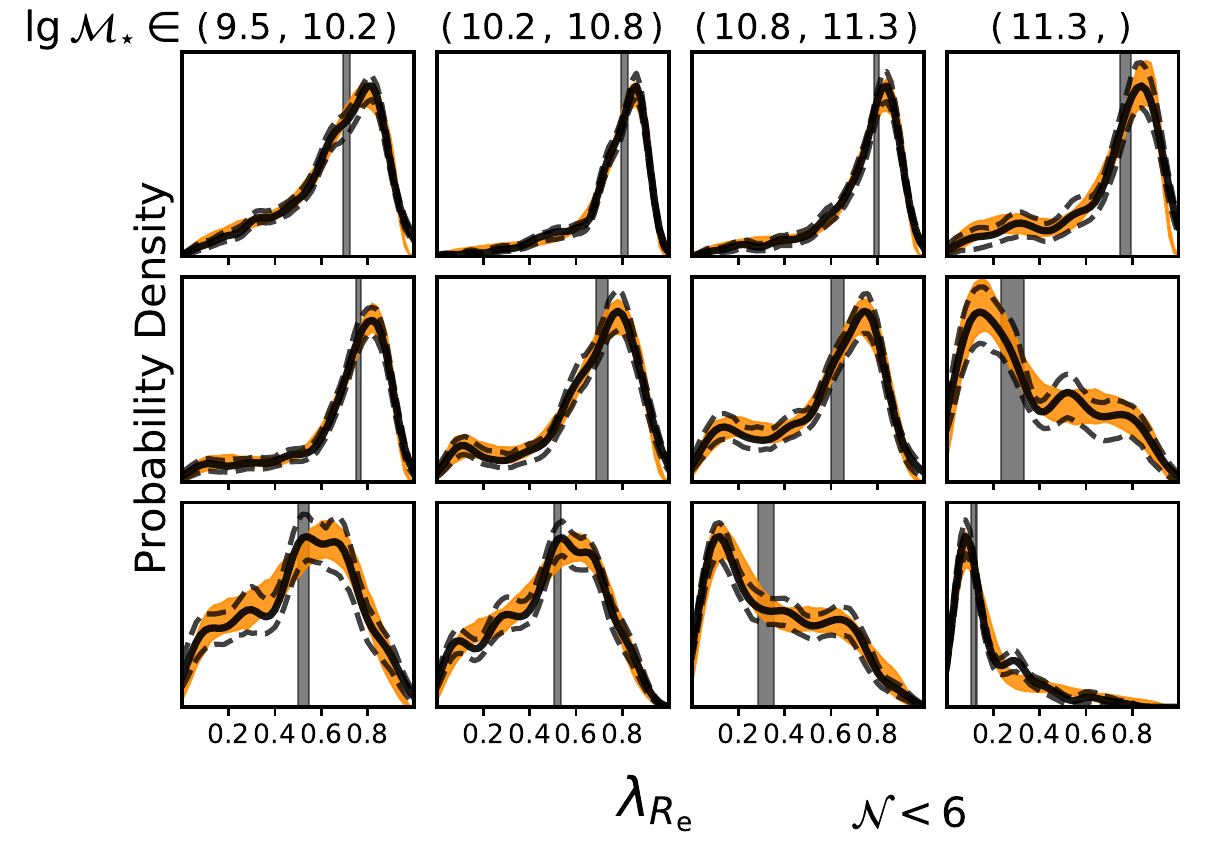}

	\vspace{+0.2cm}

	\includegraphics[width=0.47\textwidth]{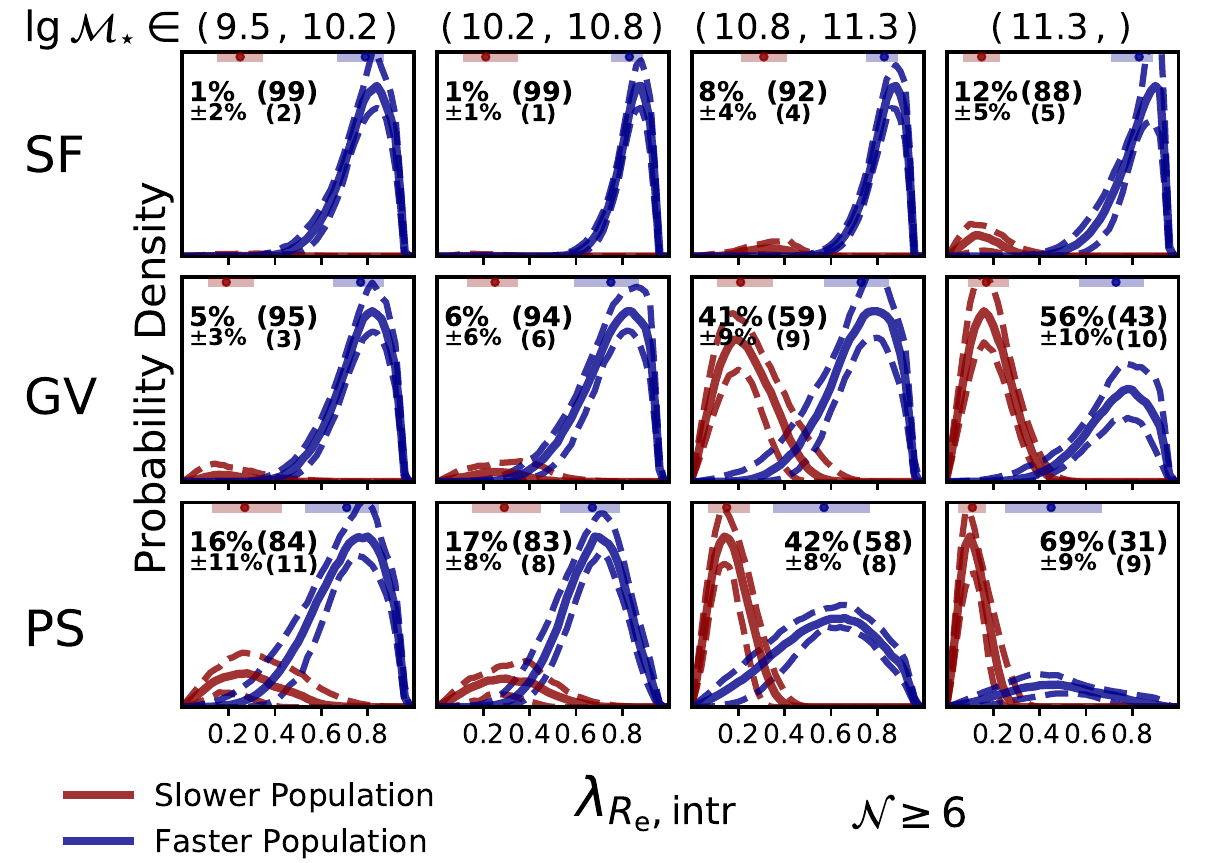}
	\includegraphics[width=0.47\textwidth]{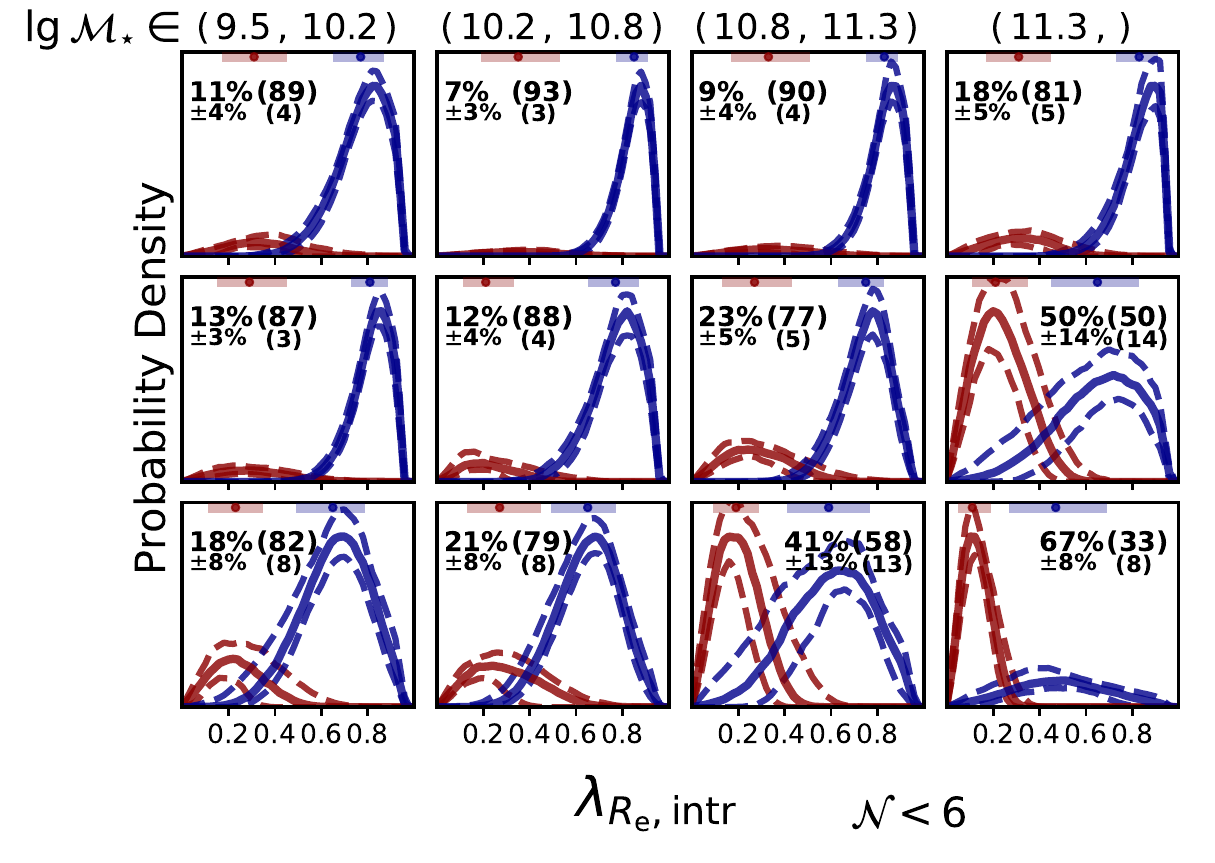}
	\caption{\textbf{The $\lambda_{R_{\rm e}}$ and $\lambda_{R_{\rm e},\mathrm{intr}}$ distributions of galaxies in different environments.}
	The same as in \autoref{fig:lam} but now the galaxies are divided into those in large galaxy groups (left-hand half) and in small groups (right-hand half) based on the dedicated group catalogue for MaNGA galaxies constructed by ref.\citep{2019arXiv191005136G}.
	}
	\label{fig:N}
\end{figure*}

\begin{figure*}
	\includegraphics[width=0.98\textwidth]{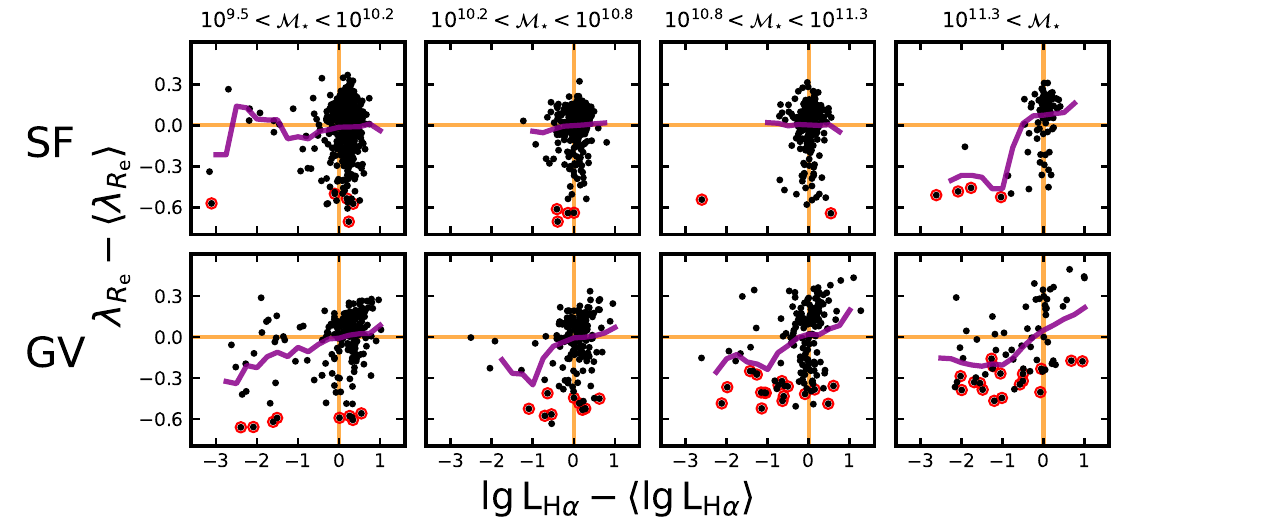}
	\caption{\textbf{The correlation between stellar angular momentum and recent star formation history.}
	The dimension-reduced version of \autoref{fig:halpha}, without LOESS smoothing. Each panel directly shows $\lambda_{R_{\rm e}} - \langle\lambda_{R_{\rm e}}\rangle$ versus a measure of recent SFH, lg$\,\mathrm{L}_{\mathrm{H}\alpha}-\langle\mathrm{lg}\,\mathrm{L}_{\mathrm{H}\alpha}\rangle$.
	$\langle\mathrm{lg}\,\mathrm{L}_{\mathrm{H}\alpha}\rangle$ is the expected H$\alpha$ luminosity given the SED SFR, according to the quadratic fit to all galaxies shown by the black dashed curve in the upper rightmost panel of \autoref{fig:halpha}.
	The magenta lines are running average to indicate the trend.
	Red circles mark the classic slow rotator galaxies defined by equation 19 of ref. \citep{2016ARA&A..54..597C}.
	}
	\label{fig:halpha2}
\end{figure*}

% Don't change these lines
\bsp    % typesetting comment
\label{lastpage}
\end{document}